\documentclass[aps,prl,twocolumn,floats]{revtex4}
\usepackage{epsfig}
\usepackage{amsmath}
\newcommand{\be}{\begin{equation}}
\newcommand{\ee}{\end{equation}}
\newcommand{\bea}{\begin{eqnarray}}
\newcommand{\eea}{\end{eqnarray}}
\newcommand{\ba}{\begin{array}}
\newcommand{\ea}{\end{array}}
\newcommand{\nn}{\nonumber}
\newcommand{\Del}{\Delta}
\newcommand{\Gam}{\Gamma}

\newcommand{\al}{\alpha}

\newcommand{\eps}{\epsilon}

\begin{document}

\title{\bf Statistical Distribution of Coulomb Blockade Peak Heights
in Adiabatically Pumped Quantum Dots}

\author{M. Blaauboer and E.J. Heller} 

\affiliation{
Department of Physics, Harvard University, Cambridge, Massachusetts 02138}
\date{\today}

\begin{abstract}
We study adiabatic quantum pumping in the resonant tunneling regime of a 
nearly-closed quantum dot, which is coupled to two leads via tunneling
barriers. Using small cyclic variations of the tunneling rates of the 
barriers as the pumping mechanism, a current is obtained which depends sensitively 
on the system parameters and exhibits peaks due to Coulomb blockade. 
The distribution of the peak heights is found for 
temperatures $\Gam \ll k_{B} T \ll \Delta$, with $\Gam$ the total decay 
width into the leads and $\Delta$ the
single-particle level spacing of the dot, and their average height is predicted 
to increase by a factor of $\frac{5\sqrt{2}}{18}\,\pi \approx 1.2$ 
upon breaking time-reversal symmetry. This is the 
pumping analog of the statistical theory of Coulomb blockade conductance peaks.
\end{abstract}

\maketitle

In recent years, the phenomenon of adiabatic quantum pumping in quantum dots
has attracted a lot of theoretical and experimental attention. It is the
generation of a d.c. current in the absence of a bias voltage by periodic 
modulations of two or more parameters, such as the shape of the dot or a 
magnetic field, that modify the quantum-mechanical 
properties of the system. The idea of adiabatic pumping was pioneered by 
Thouless \cite{thou83} for electrons at zero temperature in an isolated 
one-dimensional periodic potential. Using cyclic variations of the potential,
electrons can be pumped one by one through the system. Following early 
realizations of electron pumps based on "classical" sequential pumping 
mechanisms \cite{kouw91,classical}, 
attention is now focussed on adiabatic quantum pumping in 
quantum dots \cite{brou98,alei98,swit99,levi00}. A quantum dot is a small 
metallic or semiconducting island, confined by gates and connected to 
electron reservoirs (leads) through quantum point contacts \cite{kouw97}. 
Almost all investigations of quantum pumping to date, 
including the first experimental observation \cite{swit99}, involved open 
quantum dots. Only very recently, 
quantum pumping in closed dots has for the first time been considered.
In Ref.~\cite{levi00}, Levinson {\it et al.} study adiabatic pumping 
mediated by resonant transmission through a quantum dot 
which is isolated from the leads through tunneling point contacts. They 
predict a quantized pumped current at zero temperature, if the two tunneling
barriers are varied in such a way that the loop which describes the pumping 
cycle in parameter space encircles the entire resonance line.

Here we also consider adiabatic quantum pumping via resonant transmission
through a nearly-closed quantum dot at low temperatures, but in a new 
regime, employing small 
periodic changes in the dot potential. This pumping mechanism differs 
from Ref.~\cite{levi00} in two significant ways: firstly, the tunneling 
rates are only slightly perturbed around a point of the resonance line, 
and as a result the pumped current is not quantized. Secondly, 
for such small modulations of the potential (linear response), which
correspond to small perturbations of the energy level spectrum of the dot,
the pumped current exhibits Coulomb blockade oscillations as a function
of the Fermi energy of the dot. These are of the same origin as the well-known
fluctuations in the conductance through a quantum dot in the Coulomb blockade 
regime \cite{jala92,chan96}:
they result from the spatial structure of the quasi-bound wavefunction of the dot 
close to the leads and reflect how well this wavefunction couples to the 
electron states in the leads. We calculate the statistical distribution of 
the pumped current peak heights both in presence and absence of time-reversal 
symmetry (TRS). This leads to the prediction
of an experimentally observable increase of the average pumped peak height 
upon breaking TRS.

\begin{figure}
\centerline{\epsfig{figure=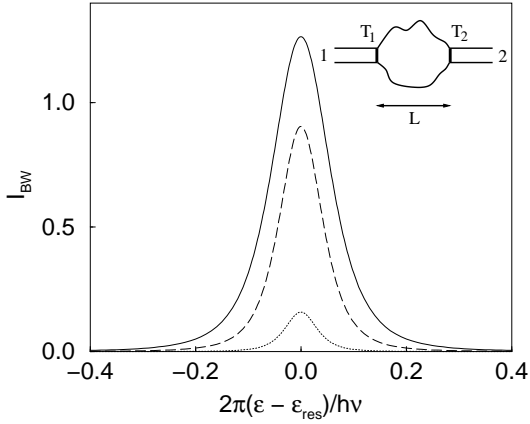,width=0.8\hsize}}
\caption[]{Pumped current (\ref{eq:currBW}) in units of $\frac{\omega e}{4\pi}\sin \phi\, 
\delta V_1\, \delta V_2$ through a nearly-closed quantum dot for $T_1$=$T_2$=0.1 (solid line), 
$T_1$=$0.1$, $T_2$=$0.05$ (dashed line), and $T_1=0.1$, $T_2=0.01$
(dotted line). Inset: Schematic picture of a quantum dot connected to two leads via
tunneling barriers.}
\label{fig:cont}
\end{figure}

Consider a quantum dot that is weakly-coupled to two single-mode leads via tunneling
point contacts with transmission probabilities  
$T_{1}$ and $T_{2}$, see Fig.~\ref{fig:cont}. If the strengths of the tunneling barriers 
are varied periodically as $V_{1}(t) = \overline{V}_{1} + \delta V_{1} \sin(\omega t)$ and
$V_{2}(t) = \overline{V}_{2} + \delta V_{2} \sin(\omega t + \phi)$, with $\delta V_{1,2}
\ll \overline{V}_{1,2}$, a current is pumped into lead 1 which is given by, 
at zero temperature \cite{brou98},
\begin{subequations}
\bea
I_{1} & = & \frac{\omega e}{2 \pi^2}\int_{A} dV_{1} dV_{2}\, \mbox{\rm Im} 
\left( \frac{\partial
s_{11}^{*}}{\partial V_{1}} \frac{\partial s_{11}}{\partial V_{2}} +
\frac{\partial s_{12}^{*}}{\partial V_{1}} \frac{\partial s_{12}}{\partial V_{2}}
\right)\label{eq:curr11} \\
& \approx & \frac{\omega e}{2 \pi}\, \sin \phi\, \delta V_1\,\delta V_2\, 
\mbox{\rm Im} 
\left( \frac{\partial
s_{11}^{*}}{\partial V_{1}} \frac{\partial s_{11}}{\partial V_{2}} +
\frac{\partial s_{12}^{*}}{\partial V_{1}} \frac{\partial s_{12}}{\partial V_{2}}
\right)\label{eq:curr12}  
\eea
\label{eq:curr1}
\end{subequations}
(and $I_{2} = - I_{1}$). Eq.~(\ref{eq:curr1}) is valid for slow perturbations such that
$\omega \ll \tau_{\rm dwell}^{-1}$, with $\tau_{\rm dwell}$ the dwell time of
particles in the dot, and $A$ denotes the area that is enclosed in parameter space 
($V_{1}$,$V_{2}$) during one period $\tau \equiv 2\pi/\omega$.
It relates the pumped current to the scattering matrix elements $s_{11}$ and
$s_{12}$ of the dot, where $s_{\alpha \beta}$ denotes the scattering amplitude 
of electrons from lead $\beta$ to lead $\alpha$. The second equation~(\ref{eq:curr12}) 
is applicable for bilinear response to the perturbations $\delta V_{1}$ and $\delta V_{2}$,
in which case the integral in (\ref{eq:curr11}) becomes independent of the pumping 
contour \cite{brou98}. This is the situation we consider here. In the low temperature regime
$k_{B}T < \Gam \ll \Delta$ transport occurs through a single bound state of the dot \cite{kouw97},
which can then be modeled as a one-dimensional system with 
$\delta$-function barriers $V_{1}\, \delta(x)$ and $V_{2}\, \delta(x-L)$ \cite{levi00,ston85}. 
The scattering matrix elements for this system are given by \cite{levi00}
\bea
s_{11} & = &  r_{1} + \frac{r_{2} t_{1}^2 e^{2ikL}}{1 - r_{1} r_{2} e^{2ikL}}
\\
s_{12} & = & s_{21} = \frac{t_{1} t_{2} e^{ikL}}{1 - r_{1} r_{2} e^{2ikL}},
\eea
with $k\equiv \sqrt{\frac{2m}{\hbar}(\eps_F-\eps)}$, $\eps$ the energy measured 
from the Fermi energy $\eps_{F}$ and $s_{22} = s_{11} (r_{1} \leftrightarrow r_{2}, 
t_{1} \rightarrow t_{2})$.
Here $r_m$ and $t_m$ denote the reflection and transmission
amplitudes of barrier $m$ [$m$=1,2],
\bea
r_{m} & = & \frac{V_{m}}{i - V_{m}} \equiv \sqrt{R_{m}}\, e^{i \phi_{m}} 
\\
t_{m} & = & \frac{i}{i - V_{m}} \equiv i \sqrt{T_{m}}\, e^{i \phi_{m}} \\
\phi_{m} & \equiv & - \arctan V_{m} - \pi/2,
\eea
with $R_{m} \equiv |r_{m}|^2$, and $T_{m} \equiv
|t_{m}|^2$. The integrand of Eq.~(\ref{eq:curr11}) can then be written as
\bea 
\Pi(T_1,T_2) & \equiv & \mbox{\rm Im} \left( \frac{\partial
s_{11}^{*}}{\partial V_{1}} \frac{\partial s_{11}}{\partial V_{2}} +
\frac{\partial s_{12}^{*}}{\partial V_{1}} \frac{\partial s_{12}}{\partial V_{2}}
\right) \nn \\
& = & \frac{(T_{1} T_{2})^{\frac{3}{2}}}{N} \left[ (\sqrt{R_1 T_2} + \sqrt{R_2 T_1}) 
(1 - \sqrt{R_1 R_2}\cdot \right. \nn \\ 
& & \left. \cos \Theta) - (1 - R_1 R_2 + \sqrt{R_1 T_1 R_2 T_2})\sin \Theta \right],
\label{eq:emis}
\eea
with
\bea
N & \equiv & \left[ (1 - \sqrt{R_1 R_2})^2 + 2 \sqrt{R_1 R_2} (1 - \cos \Theta) 
\right]^2 
\label{eq:noemer}\\
\Theta & \equiv & \phi_{1} + \phi_{2} + 2kL. \nn
\eea

\begin{figure}
\centerline{\epsfig{figure=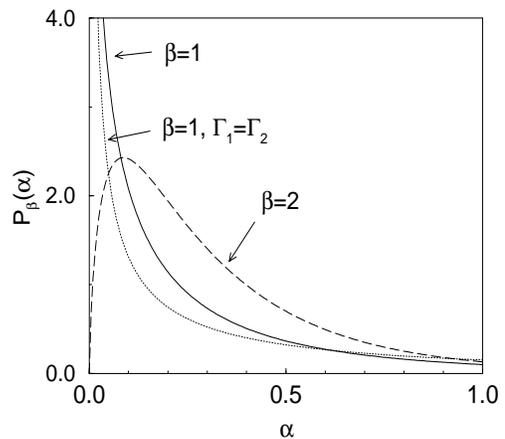,width=6.5cm,height=6cm}}
\caption[]{Distribution $P_{\beta}(\al)$ of pumped current peak heights 
for symmetric leads in
zero magnetic field ($P_{1,\Gam_1=\Gam_2}(\al)$, dotted line), asymmetric leads in 
zero magnetic field ($P_1(\al)$, solid line) and asymmetric leads in a magnetic field
($P_2(\al)$, dashed line).
}
\label{fig:dist}
\end{figure}

For a nearly-closed quantum dot, the transmission probabilities 
$T_{1},T_{2} \ll 1$. If $\delta V_{m}$ $\ll$ $V_{m}$  and hence 
$\delta T_{m}$ $\ll$ $T_{m}$, for $m \in \{1,2\}$, the quantum dot remains 
in the Coulomb blockade regime during the entire pumping cycle. $\Pi$ then 
exhibits sharp resonances when $N=0$. For a completely-closed dot 
($T_1 = T_2 = 0$)
the resonance condition is given by $\cos \Theta = 1$. In case of a nearly-closed dot the 
cosine can be expanded around this resonance as $1 - \cos \Theta
\approx \frac{1}{2} \left( \frac{d\Theta}{d\eps} \right)^2 (\eps - \eps_{\rm res})^2$. Up
to lowest order in $T_1$ and $T_2$ \cite{exact} the pumped current $I_{BW}$ is then given by
\bea
I_{BW} = \frac{\omega e}{4 \pi} \sin \phi\, \delta V_1 \delta V_2\, 
\frac{(T_{1}\,T_{2})^{\frac{3}{2}} (T_1^{\frac{1}{2}} + T_2^{\frac{1}{2}}) (T_1 + T_2)}{\left[ 
\left( \frac{T_1 + T_2}{2} \right)^2 + \left( \frac{\eps - \eps_{\rm res}}{\hbar \nu} 
\right)^2 \right]^2},
\label{eq:currBW}
\eea
with $\hbar \nu \equiv d\eps /d \Theta$ and $\nu$ the attempt frequency, 
the inverse of the round-trip travel time between the two barriers. 
Equation (\ref{eq:currBW}) is valid for $k_B T$ $\ll$ $\Gam$ $\ll$ $\Del$, 
$|\eps - \eps_{\rm res}| \ll \hbar \nu$, and bilinear response to the perturbations
$\delta V_1$ and $\delta V_{2}$. Figure~\ref{fig:cont} shows that $I_{BW}$ is sharply 
peaked around the resonant level $\eps_{\rm res}$ \cite{wei00}. The effect of finite 
temperatures can be incorporated by thermally averaging
(\ref{eq:currBW}) as $I \equiv - \int d\eps\, I_{BW} f^{'}(\eps,T) \approx 
\frac{1}{4k_{B} T} \int_{\eps_{\rm res} - \frac{1}{2}\hbar \nu (T_1 + T_2)}^{\eps_{\rm res}}
d\eps\, I_{BW}$, where $f(\eps,T)$ $\equiv$ $[1 + {\rm exp}(\eps/k_{B}T)]^{-1}$ denotes
the Fermi function. This results in the peak heights, for $\Gam \ll k_{B}T \ll \Del$,
\bea
I_{\rm max} = \frac{\omega e (2 + \pi) \bar{\Gamma}^{3/2}}{16 \pi k_{B} T 
\sqrt{\hbar \nu}}\, \sin \phi\, \delta V_1 \delta V_2\, \alpha.
\label{eq:curr2} 
\eea
Here $\alpha$ denotes the amplitude of the current normalized by the mean resonance 
width $\bar{\Gamma}$,
\bea
\alpha \equiv \frac{(\Gam_{1}\, \Gam_{2})^{\frac{3}{2}}\, (\Gam_1^{\frac{1}{2}} + 
\Gam_2^{\frac{1}{2}})}{\bar{\Gam}^{\frac{3}{2}}\, (\Gam_1 + \Gam_2)^2},
\label{eq:heights}
\eea
with $\Gam_{1,2} \equiv \hbar \nu T_{1,2}$. 
Equations~(\ref{eq:currBW}) and (\ref{eq:curr2}) are the
pumping analogs of, resp., the Breit-Wigner conductance $G_{\rm BW}$ 
through a quantum dot in the resonant tunneling regime \cite{ston85} and its peak heights 
$G_{\rm max}$ \cite{been91},
\bea
G_{BW} & = & \frac{2e^2}{h}\, \frac{\Gam_{1}\Gam_{2}}{\frac{1}{4} \Gam^2 + (\eps - 
\eps_{\rm res})^2}, \hspace*{0.6cm} T=0,\, \Gam \ll \Del.
\label{eq:BW}
\\
G_{\rm max} & = & \frac{e^2}{h}\, \frac{\pi}{4k_{B} T}\, \frac{\Gam_{1}\Gam_{2}}
{\Gam_{1} + \Gam_{2}}, \hspace*{1.1cm} \mbox{\rm $\Gam \ll
k_{B}T \ll \Del$}.
\label{eq:maxpeak}
\eea
Jalabert {\it et al.}\cite{jala92} calculated the statistical distributions 
of the Coulomb blockade conductance peaks (\ref{eq:maxpeak}) for single-mode 
leads in presence and 
absence of time-reversal symmetry, which were in excellent 
agreement with subsequent experiments \cite{chan96}. In an analogous manner we 
now proceed to calculate the distribution of the
pumped current peaks (\ref{eq:heights}). Assuming equivalent single-mode point contacts
with energy-independent mean tunneling rates $\bar{\Gam}_1$ $=$ $\bar{\Gam}_2$ $\equiv$
$\bar{\Gam}/2$ (valid if the separation between the barriers $L \gg \lambda_F$, with 
$\lambda_F$ the Fermi wavelength), the distributions of $\Gam_1$ and $\Gam_2$ for 
a chaotic quantum dot are given by the Porter-Thomas distribution \cite{jala92} 
\bea
P_{\beta}(\Gam_m) = \left( \frac{\beta}{2 \bar{\Gam}} \right)^{\beta/2} 
\frac{1}{G(\beta/2)}\, \Gam_m^{\beta/2 - 1} e^{-\beta \Gam_m/ 2 \bar{\Gam}},
\label{eq:PT}  
\eea
[$m=1,2$], where $G$ denotes the Gamma function  and the symmetry index $\beta=1(2)$ 
in the presence (absence) of time-reversal symmetry. For independently 
fluctuating tunneling rates the distribution of $\alpha$ is then given by
$P_{\beta}(\alpha) = \int d\Gam_1 \int d\Gam_2 P_{\beta}(\Gam_1) P_{\beta}(\Gam_2)
\delta (\alpha - \frac{(\Gam_{1}\, \Gam_{2})^{\frac{3}{2}}\, (\sqrt{\Gam_1} + 
\sqrt{\Gam_2})}{\overline{\Gam}^{\frac{3}{2}}\, (\Gam_1 + \Gam_2)^2})$. 
In case of a symmetric dot with 
$\Gam_{1} = \Gam_{2}$ in zero magnetic field this yields
\bea
P_{1,\Gam_1 = \Gam_2} (\al) = \frac{4}{3 \sqrt{2 \pi}\, (2 \al)^{2/3}}\, e^{- \frac{1}{2} 
(2\al)^{2/3}}.
\label{eq:distsymm}
\eea
For the more general case of an asymmetric dot with $\Gam_1\neq \Gam_2$, it is convenient 
to write (\ref{eq:heights}) as 
$\al$ $\equiv$ $\al_{1} + \al_{2}$ $=$ $\frac{\Gam_{1}^{2}\, \Gam_{2}^{\frac{3}{2}}}
{\overline{\Gam}^{\frac{3}{2}}\, (\Gam_1 + 
\Gam_2)^2} + \frac{\Gam_{1}^{\frac{3}{2}}\, \Gam_{2}^{2}}{\overline{\Gam}^{\frac{3}{2}}\, 
(\Gam_1 + \Gam_2)^2}$, calculate the distributions
$P_{\beta}(\al_1)$ and $P_{\beta}(\al_2)$ and from these
$P_{\beta}(\al) = \int d\al_1\, \int d\al_2\, P_{\beta}(\al_1) P_{\beta}(\al_2) 
\delta(\alpha - (\alpha_1 + \alpha_2))$. This results in the following expressions (exact for
$P_1(\al)$ and numerical approximation to within 0.1 \% accuracy for $P_2(\al)$)
\bea
P_{1}(\al) = \frac{C_1}{\al^{\frac{1}{3}}}\, 
\int_{0}^{1} dx\, \frac{1}{\left[ x(1-x) \right]^{\frac{2}{3}}} 
e^{-\frac{1}{2} \left( \frac{35}{3} \al \right)^{\frac{2}{3}} \left[x^{\frac{2}{3}} + 
(1-x)^{\frac{2}{3}} \right]} 
\label{eq:GOE}
\eea
\bea
P_{2}(\al) =  C_2 \al^{\frac{4}{5}} \int_{0}^{1} dx\, \frac{1}{\left[ x(1-x) 
\right]^{\frac{1}{10}}}\, 
e^{- \left( \frac{35}{3} \al \right)^{\frac{2}{3}} \left[x^{\frac{2}{3}} + 
(1-x)^{\frac{2}{3}} \right]},
\label{eq:GUE}
\eea
with $C_1 \equiv 2 (35/3)^{\frac{2}{3}}/9\pi$ and $C_2 \approx 46.6$. 
The distributions $P_{1}(\al)$ and $P_{2}(\al)$ are plotted in Fig.~\ref{fig:dist}. 
Breaking TRS increases the mean amplitude from 
$\overline{\al}_{B=0}$ $=$ $\frac{12}{35} \sqrt{\frac{2}{\pi}}$
$\approx$ 0.27 to $\overline{\al}_{B\neq 0}$ $=$ $\frac{4}{21} \sqrt{\pi}$ $\approx$ 0.34. 
Simultaneously, the variance of $\al$ decreases from $\sigma^2(\al)_{B=0} = \frac{15}{64} 
- \left(\frac{12}{35} \right)^2 \frac{2}{\pi} \approx 0.16$ 
to $\sigma^2(\al)_{B\neq 0} = \frac{6}{35} - \left( \frac{4}{21} \right)^2 \pi 
\approx 0.06$ \cite{open}. The distributions (\ref{eq:GOE}) and (\ref{eq:GUE}) are similar to the 
ones for conductance in Ref.~\cite{jala92}, with means and variances being of the same 
order of magnitude. Jalabert {\it et al.} tested 
their prediction of the Porter-Thomas distribution (\ref{eq:PT}) numerically and obtained 
excellent agreement. This analysis also serves as a numerical test of the 
validity of $P_1(\al)$ and $P_2(\al)$, since these follow analytically
from (\ref{eq:PT}).

The above model is entirely based on a noninteracting particle picture. Since
the charge distribution in a quantum dot is modified by interactions among the 
electrons \cite{evan93}, the question arises
whether one may neglect the effect of e-e interactions on quantum pumping in the 
Coulomb blockade regime here. Measurements of the phase of the reflection coefficient 
of a quantum dot in the linear quantum Hall regime were observed to be in
good agreement with a noninteracting model if the dot was close 
to resonance, with deviations occurring further away from the resonance 
condition \cite{buks96}. Therefore the present noninteracting model is expected to 
give, at least in first approximation, a good description  of quantum pumping (and
the pumped peak heights) in a nearly-closed dot close to resonance. 
Studying the effects of e-e interactions in this regime forms an interesting 
direction for future research.

The change in $\overline{\al}$ upon application of a magnetic field can be experimentally observed 
in small strongly-pinched quantum dots  with typical diameter $\sim 1\, \mu$m, Fermi wavelength
$\lambda_{F} \sim 10\,$ nm, level spacing $\Delta \sim 10\, \mu$eV and barrier conductances 
$G_{1,2} \ll \frac{2e^2}{h}$ \cite{folk00}. Adiabatic quantum pumping through the dot may 
be achieved at low temperatures $T \sim 10$ mK, for which $\Gam < k_{B} T < \Del$, 
by periodically modulating either the tunneling barriers with two small-amplitude 
phase-shifted rf signals \cite{kouw91}, or the confining potential of 
the dot by two shape-changing voltages. 
Measuring the pumped current peaks as a function of an additional split-gate voltage,
which varies the electron density of the dot and hence the Fermi energy $\eps_{F}$, 
would then allow to assemble their statistics. The modulation amplitudes must be small 
enough for the perturbation of the energy levels of the dot to be $\ll \Delta$ for each
value of $\eps_{F}$. Another important requirement is that the dephasing 
(loss of phase coherence) and leakage (escape of electrons through the leads) times 
must be larger than the resonant tunneling time. The former have been measured for 
nearly-closed quantum dots
at temperatures $\sim 10$ mK in Ref.~\cite{folk00} and are given by $\tau_{\rm dephasing} 
\sim 1$ ns and 
$\tau_{\rm escape} \sim 0.1$ ns, respectively. Since the typical dwell time for electrons
on the dot $\tau_{\rm dwell} \sim 10^{-12} - 10^{-11}$ s for $G_{1,2} \sim 0.01-0.1\, e^2/h$,
we expect that quantum pumping should not be strongly affected by these inelastic processes. 
For typical 
pumping frequencies $f\sim 10$ MHz \cite{kouw91,swit99}, also the adiabaticity condition
$ \omega =2\pi f \ll \tau_{\rm dwell}^{-1}$ is fullfilled.

Statistical theories based on RMT such as 
the above and in Ref.~\cite{jala92} do not account for correlations between adjacent
peaks. In measurements of the conductance, however, these 
were observed \cite{chan96,cron97} and semiclassical analysis suggested that they 
result from short-time dynamics in 
the dot \cite{nari99}. An interesting direction for future research is therefore to study the
correlations of pumped current peak heights. Another open question concerns 
the effects of dephasing 
on these peak heights. For conductance, 
the normalized change in average peak heights upon breaking TRS in the absence 
of dephasing $\delta g \equiv (\langle g \rangle_{B\neq 0} - \langle g \rangle_{B=0})/
\langle g \rangle_{B\neq 0}$ is independent of temperature \cite{alha982}. $\delta g$
can thus be used to investigate the temperature dependence of the dephasing time, 
$\tau_{\phi}$, since any observed temperature dependence in measurements of 
$\delta g$ is due to $\tau_{\phi}$ \cite{folk00}. Similarly the change in average
pumped current peak heights may provide a useful tool to study dephasing.

In conclusion, we shave studied adiabatic quantum pumping and its statistics in the 
coherent resonant tunneling regime of a nearly-closed quantum dot. Using random-matrix theory, 
the average peak height of the pumped current is predicted to display an observable increase 
in the presence of a symmetry-breaking magnetic field.

Stimulating discussions with Y. Levinson and C.M. Marcus are gratefully acknowledged. 
This work was supported by the Netherlands Organisation for Scientific Research (NWO), 
by NFS grant CHE-0073544 and by an NSF MRSEC grant.


\begin{thebibliography}{99}
\bibitem{classical}
The word "classical" here refers to the fact that these pumping mechanisms
do not involve quantum interference, but are based on sequential tunneling 
processes in the Coulomb blockade regime of a quantum dot and thereby resemble 
classical peristaltic pumping.
\bibitem{exact}
The complete expanded expression for $\Pi(T_1,T_2)$ is given by 
$\Pi(T_1, T_2)$ $=$ $(T_{1}\, T_{2})^{3/2}$ $\left[ 
\frac{1}{2} \right.$  $(\sqrt{T_1} + \sqrt{T_2})$ $(T_{1} + T_{2}) $ $-$ $(T_{1} + T_{2} + 
\sqrt{T_{1} T_{2}})$ $\left( \frac{\eps - \eps_{\rm res}}{\hbar \nu} \right)$ $+$ $  
\frac{1}{2}$ $(\sqrt{T_1} + \sqrt{T_2})$ $\left. \left( \frac{\eps - \eps_{\rm res}}{\hbar \nu} 
\right)^2 \right]$ $/$ $\left[ \left( \frac{T_1 + T_2}{2} \right)^2 \right.$ $+$ $\left. 
\left( \frac{\eps - \eps_{\rm res}}{\hbar \nu} \right)^2 \right]^2$.
\bibitem{open}
A similar decrease in variance has been observed in open quantum dots, see 
Ref.~\cite{swit99}.
\end{thebibliography}
\end{document}